\documentclass[twocolumn,showpacs,preprintnumbers,amsmath,amssymb,endfloats]{revtex4}

\newcommand{\br}[0]{{\bf r}}
\newcommand{\brp}[0]{{\bf r}^{\prime}}
\newcommand{\brb}[0]{({\bf r)}}
\newcommand{\brbp}[0]{({\bf r^{\prime})}}
\newcommand{\balpha}[0]{\boldsymbol{\alpha}}
\newcommand{\bbeta}[0]{\boldsymbol{\beta}}
\newcommand{\bsigma}[0]{\boldsymbol{\sigma}}
\newcommand{\rha}[0]{{\bf \hat{r}}}
\newcommand{\bk}[0]{{\bf k}}

\newcommand{\Mel}[3]{\langle #1 | #2 | #3 \rangle}

\begin{document}

\draft
\preprint{}


\title{Systematic theoretical study of the spin and orbital
magnetic moments \\ of 4d and 5d interfaces with Fe films}

\author{R. Tyer, G. van der Laan, W. M. Temmerman, and Z. Szotek}
\address{Daresbury Laboratory, Warrington WA4 4AD, UK}
\author{H. Ebert}
\address{University of Munich, D-81377 Munich, Germany}
\date{\today}


\begin{abstract}

Results of {\it ab initio} calculations using the relativistic 
Local Spin Density theory  are presented for the magnetic moments of 
periodic $5d$ and $4d$ transition metal interfaces with bcc Fe(001).  
In this systematic study we calculated the layer-resolved spin and
orbital magnetic moments  over the entire series.  For the
Fe/W(001) system, the Fe spin moment is reduced whilst its orbital
moment is strongly enhanced. In the W layers a spin moment is induced,
which is antiparallel to that of Fe in the first and fourth W layers but parallel
to Fe in the second and third W layers.  The W orbital moment does not follow
the spin moment. It is aligned antiparallel to Fe in the first two W
layers and changes sign in the third and fourth W layers. 
Therefore, Hund's third rule is violated in the first and third W layers,
but not in the second and fourth W layers.
The trend in the spin and orbital moments over the $4d$ and $5d$
series  for multilayers is quite similar to previous impurity
calculations.  These observations strongly suggest that these effects
can be seen as a consequence of the hybridization between $5d$ ($4d$) and Fe
which is mostly due to band filling, and to a lesser extent geometrical
effects of either single impurity or interface.

\end{abstract}

\maketitle


\section{Introduction}

Driven by recent technological interests in magnetic recording and
data storage, it has become possible to grow well-characterized  thin
films, multilayers, and nanostructures. Simultaneously, advances in
synchrotron radiation instrumentation have made it possible to 
determine the electronic and magnetic properties of these systems in an
element-specific way.   X-ray magnetic circular dichroism (XMCD) in
conjunction with the sum rules enables
the separation of the spin and orbital contributions to the total
magnetic moments \cite{sumrules} while complementary soft x-ray resonant
magnetic scattering measurements can provide details about the layer
dependence of the magnetic moments, periodicity of the magnetic domain
structures, and roughness of the  magnetic layers
\cite{Durr,Seve,Jaouen}.   
Developments of theoretical
models which treat these magnetic systems have also been quite
successful.

It has been known already for some time that the
spin and orbital magnetic moments at surfaces are
often enhanced compared to their bulk values due to symmetry
breaking and $d$ band narrowing  at the surface
\cite{Krakauer,NiLz,Hjortstam}.
Electron hybridization at interfaces can give rise to charge transfer
across the interface \cite{Dhesi} resulting in a change of the
density of states (DOS) near the Fermi level. 
This can enhance the spin and orbital magnetic moments
in magnetic layers and can induce magnetization in adjacent
``non-magnetic'' layers. In giant magnetoresistance (GMR) materials,
such as Co/Cu multilayers, the magnetic layers are
(anti)ferromagnetically coupled by the Ruderman-Kittel-Kasuya-Yosida
(RKKY) interaction, which is accompanied by an induced oscillatory magnetic moment in the
``non-magnetic'' spacer layer. The magnetic properties of two metals
near their common interface are essentially determined by the differences in
electronegativity. Similarly to the model of Friedel \cite{Friedel} the
excess nuclear charge displaces locally the mobile electrons, i.e.\ the
electrons close to the Fermi level, until the displaced charges totally screen
out the nuclear charge. This effect occurs at the interface
over distances of the order of a few interatomic distances, i.e.\ the
screening is strongly localized. In Fe with only few spin-up electrons
close to
$E_{\text{F}}$, the screening is almost uniquely due to spin-down $d$
electrons. 
 
Most of the attention so far has been focused on the magnetism in $3d$
materials, although the possibility to observe magnetism in $4d$
and $5d$ elements in a surface geometry has been noticed early on
\cite{Eriksson}.
Impurity systems have been thoroughly studied by relativistic
calculations and in particular the systematics of ${\mathsf{3d}}$,
${\mathsf{4d}}$, and
${\mathsf{5d}}$ impurities in a magnetic host have been established
\cite{Ebert,Popescu}. 
Experimentally, sizeable spin and orbital magnetic moments in $5d$ 
transition metals due to the hybridization with a magnetic ${\mathsf{3d}}$ element were
found in e.g.\  FePd \cite{Kamp} and CoPt \cite{Grange} alloys, Ni/Pt 
multilayers \cite{WilhelmNiPt}, and Ir impurities
\cite{Krishnamurthy}.
Also Fe mono- and bi-layers on W(110) received a considerable theoretical
interest
\cite{Qian1999,Qian2001,Galanakis2000,GalanakisBis}.

Recently, the spin and orbital moments for W/Fe, Ir/Fe
multilayers were reported in Ref.~\cite{Wilhelm}. Using atomistic
arguments the authors concluded solely from the measurements of these
two ${\mathsf{5d}}$ elements that the systematic behavior across the
series was different than for ${\mathsf{5d}}$ impurities.  Our
systematic study across the entire ${\mathsf{5d}}$ series shows that in
the vicinity of W and Ir a sign change of the orbital moment occurs.
Therefore, it should come as no surprise that the orbital moment of W
can be positive or negative at the interface or for the impurity,
respectively.

By contrasting the results of ${\mathsf{5d}}$ impurities with
${\mathsf{5d}}$ interfaces in Fe, we will be able to establish the role
of band filling versus the importance of geometry or structure of both the
impurity and interface.
This is an important issue since the work in Ref.~\cite{Wilhelm} implied
that the induced magnetic behavior of ${\mathsf{5d}}$ layers may be
radically different than that of impurities and alloys. If this were to be
true then we would expect the magnetic properties to be determined more or less by 
structural considerations only and not to see any commonality between
impurity and interface as a function of atomic number Z. In the case
of an W impurity in Fe, the magnetic properties of the former are determined 
by $5d$-$3d$ hybridization, whilst at a W 
interface in Fe this $5d$-$3d$ hybridization is
reduced by the in-plane $5d$-$5d$ hybridization.
A question arises as to whether this ${\mathsf{5d}}$-${\mathsf{3d}}$ hybridization
is sufficiently reduced to alter, drastically enough,  
the magnetic properties of W interfaces in Fe in comparison to W impurities
in Fe to give rise to the proposed radical new behavior as suggested in
Ref.~\cite{Wilhelm}.

In the present study, the systematics of ${\mathsf{5d}}$ and
${\mathsf{4d}}$ interfaces with a Fe substrate will be considered. The
relative alignment of the spin and orbital moments will be compared
with the predictions based on Hund's third rule. 
This rule states that spin and orbital moment should be parallel
(antiparallel) for more (less) than half-filled shell, 
c.f. Fig.~\ref{Hund} where the atomic values of these quantities are shown.
Although
strictly valid only for single atoms, this rule seems to be applicable also to
solids with only a few exceptions, such as U metal \cite{Hjelm} and
vanadium in VAu$_4$ \cite{Galanakis2001a} and VPt$_3$ \cite{Galanakis2001b}. The violation reported in
Fe/W multilayers \cite{Wilhelm} seems surprising, since W as an impurity
and in alloys does not show this violation \cite{Ebert,Schutz}.
The opposite situation arises for Ir that obeys the
third Hund's rule in Fe/Ir multilayers \cite{Wilhelm} but violates it
as a 5d impurity in Fe
\cite{Ebert,Schutz,Kornherr,Tyer}.
Therefore, in this study we will compare the electronic and magnetic
properties of a ${\mathsf{5d}}$ transition metal with those of an
impurity.


\section{Calculational}

Calculations were performed within the framework of relativistic
local spin density (LSD) band theory, where analogous to the
non-relativistic case, a set of coupled Kohn-Sham-Dirac equations is
derived which describes the ground state of a relativistic many-electron
system
\cite{PS} with the corresponding Hamiltonian of the form

\begin{eqnarray}
\label{eq:relH}
H = &&-ihc\balpha \cdot \mathbf{p} + \bbeta mc^2 \nonumber \\
&& + V^{\text{eff}} [n\brb,m\brb] +
\bbeta \sigma_z B^{\text{eff}} [n\brb,m\brb] \;\;,
\end{eqnarray}

\begin{eqnarray}
V^{\text{eff}} [n\brb,m\brb] &=& V^{\text{ext}} \brb +
\frac{\delta E^{\text{XC}} [n\brb,m\brb] }{\delta n\brb} \nonumber \\
&&+\; e^2 \int \frac{n\brbp}{| \br -
\brp |} d\brp  \;\;, \\ B^{\text{eff}} [n\brb,m\brb] &=&
\frac{eh}{2mc}
\frac{\delta E^{\text{XC}} [n\brb,m\brb] }{\delta m\brb} \;\;, \\
n\brb &=& \sum \psi^+_i \brb \psi_i \brb \;\;, \\
m\brb &=& \sum \psi^+_i \brb \bbeta\sigma_z \psi_i \brb \;\;,
\end{eqnarray}

\noindent
where $\psi_i \brb$ is a four-component one-electron Dirac spinor.
The matrices $\mathbf{\alpha_i}$ and $\bbeta$ are the standard Dirac matrices
while $\bsigma$ is the vector of the 4$\times$4 Pauli matrix.

The spin-polarized relativistic LMTO method \cite{relLMTO,Solovyev,Beiden} was
used to calculate the spin and orbital magnetic moments. 
The Hamiltonian and overlap matrices corresponding to
Eq.~(\ref{eq:relH}) are expressed in terms of the basis set of muffin-tin
orbitals (MTO) which are constructed from the solutions,
$\phi_{\kappa, \mu}$, to the single-site Dirac equation, given by

\begin{equation}
\phi_{\kappa, \mu} (E,\br ) = i^l \left( \begin{array}{c}
                 g_{\kappa} (E,r) \chi_{\kappa}^{\mu} (\rha) \\
                 if_{\kappa} (E,r) \chi_{-\kappa}^{\mu} (\rha) 
                  \end{array}  \right)   \,\, ,
\end{equation}

\noindent
where $g_{\kappa} (E,r)$ and $f_{\kappa} (E,r)$ are the radial
functions and the spin-angular function is

\begin{equation}
\chi_{\kappa}^{\mu} (\rha) = \sum_{m_s = \pm \frac{1}{2}} C(l 
s j; \mu - m_s, m_s) Y_l^{\mu - m_s } (\rha) \chi_{m_s} \,\, .
\label{eq:angfn}
\end{equation}

\noindent
Here $C(l s j;m_l,m_s)$ is a
Clebsch-Gordan coefficient and $Y_l^m(\rha)$ a spherical
harmonic.
 
 
The values for the spin and orbital magnetic moments are obtained using

\begin{eqnarray}
\mu_S  &=& \mu_{\text{B}}\sum_{j,\bk} \Mel{\Psi^{j
\bk}}{\bbeta
\sigma_z}{\Psi^{j \bk}} \Theta( E_F - E_{j\bk}) \,\, , \\
\mu_L  &=& \mu_{\text{B}} \sum_{j,\bk} \Mel{\Psi^{j
\bk}}{\bbeta
l_z}{\Psi^{j \bk}} \Theta( E_F - E_{j\bk}) \,\, ,
\end{eqnarray}

\noindent
where $\Psi^{j \bk}$ and E$_{j\bk}$ are the eigenvector and
eigenvalue, respectively, of the $j$-th band at the point $\bk$ of the
Brillouin zone.

Self-consistent relativistic calculations were performed for the
electronic and magnetic properties of periodic multilayers consisting
of $n$Fe(001)/$m$(${\mathsf{5d}}$) where ${\mathsf{5d}}$ can be
Ta, W, Re, Os, Ir,
Pt, or Au. The layers were constructed to be consistent with the bcc
structure. The layer thickness $n$ of Fe took on the values $n$=3 and 5
while the layer thickness $m$ of
${\mathsf{5d}}$ has the values  $m$=1, 3, 5, and 7.

Basis sets with $l_{\text{max}}=2$ were used for both Fe and
${\mathsf{5d}}$. Convergence tests were performed to determine the
influence of the inclusion of the higher angular
momentum basis states, $l_{\text{max}}=3$, and the values of the
magnetic moments were found to be well converged with the inclusion of
the $l_{\text{max}}=3$ states changing the values of the spin and
orbital moments  typically by less than 0.01 and 0.003 $\mu_{\text{B}}$,
respectively.


The planar lattice constant of bcc Fe was used for both the Fe and
the ${\mathsf{5d}}$ layers. Perpendicular to the planes the lattice
parameter of Fe was used for the Fe layers, while for ${\mathsf{5d}}$
the lattice parameter perpendicular to the planes was determined such
that the volume around
${\mathsf{5d}}$ was equal to the experimental volume for the
elemental ${\mathsf{5d}}$ system. This procedure was used to model the
relaxation of the lattice at the interface.



\section{RESULTS AND DISCUSSION}


\subsection{Electronic and magnetic properties of 5Fe(001)/$m$W}
\label{sec:A}

In Fig.~\ref{LSthickness} the variations of the spin and orbital
moments are shown as a function of W thickness for a fixed Fe
thickness of 5 layers. The numerical values can be found in
Table~\ref{Qtab1}. Note that at the Fe side of the Fe/W interface the
spin moment is decreased in comparison with the other Fe layers, whilst
the orbital moment shows a strong enhancement in the interface layer.
On the W side of the interface, a spin moment on W is induced by the
Fe which is aligned antiparallel. This W moment shows an oscillatory
behavior, becoming aligned to the Fe spin moment in the two subsequent 
W layers and antiparallel again in the fourth layer. 
The W orbital moments are initially  
aligned antiparallel to the Fe orbital moment and become parallel
from the third W layer onwards. From these results it can be seen that
in the W interface layer the spin and orbital magnetic moments are
parallel aligned demonstrating that Hund's third rule is
violated as reported by the experimental work of
Ref.~\onlinecite{Wilhelm}. 
This can also be seen in Table~\ref{Qtab1} where for 5Fe/7W the W1 interface and W3 layers 
show a violation of Hund's third rule whereas layers W2 and W4 have spin and 
orbital magnetic moment anti-parallel aligned. 

Table~\ref{Qtab1} also relates the trends 
in spin and orbital magnetic moments to charge transfer. 
The charge transfer at the interface can be sizeable for all systems considered: 
the 5Fe/W, 5Fe/3W, 5Fe/5W and 5Fe/7W systems. 
From the distribution of the Fe and W electrons into d and sp channels, 
we see that for Fe the spin magnetic moment contribution from the sp electrons is antiparallel 
to the contribution from the d electrons. This is what one would expect from Fe. 
However, for W, these contributions are parallel and are in accordance with a 
magnetic moment induced by hybridization with Fe. Actually, the contribution of the 
sp electrons is as large as $\sim30\%$ of the total spin moment. 
One can also see that the induced spin magnetic moment is largest on the W1 interface layer 
where the hybridization with Fe is most sizeable.

The oscillations of the spin moments in the W layers are further
analysed in Fig.~\ref{Centroid} where
the centroid positions of the spin-up (majority) and spin-down
(minority) $d$ bands are presented for each of the layers of 5Fe/3W.
The reduction of the Fe spin moment at the
interface can be seen to be caused by the centroid position of the
minority spin band lowering its energy. For the W layers the centroid
positions of the spin bands lie above the Fermi energy
($E_{\text{F}}$) and a small splitting of these is induced. This
splitting between the spin-up and spin-down bands reverses sign from the
W interface layer to the next W layer.

The sign of the spin moment in the W interface layer being
aligned antiparallel to Fe is a consequence of the Fe-W hybridization
across the interface. The minority Fe $d$ bands lie energetically close
to the minority W bands and a relatively pronouned hybridization between these bands
develops. This pulls the W minority $d$ bands down in energy. In
contrast the majority Fe and W $d$ bands are energetically well
separated and substantially less hybridization occurs. 
Consequently, the W minority d bands are pulled down in energy. 
Therefore these W minority d bands will be preferentially occupied and this allows 
a W spin magnetic moment, opposite in sign to that of Fe, to develop.

Figure~\ref{DOS} further demonstrates the hybridization of Fe and W at
the interface. It shows the layer decomposed DOS
for the 5Fe/5W system. Whilst in Fe and W layers away from the
interface the DOS resembles that of bulk Fe and W,
respectively, a substantial hybridization at the interface is noted. In
particular the minority Fe $d$ band loses structure and its center of
mass is shifted to lower energy. The  majority and minority Fe $d$
bands in the energy region of $E_{\text{F}}$ and 0.2 Ry below
$E_{\text{F}}$ hybridise strongly with the W $d$ bands,
creating in the  spin-resolved W $d$ bands a similar depletion and
increase as in the Fe DOS.
This hybridization rapidly disappears in subsequent W layers.

These results are in contrast with the findings of
Ref.~\onlinecite{Ceri} where for Gd/W multilayers no
hybridization-induced spin polarization in the W layer is reported and
spin polarization is only found to occur as a consequence of expanding
the volume around the W sites sufficiently to induce a volume prompted
spin polarization. From the present study we see that hybridization of
$d$ electrons at the interface between Fe, which has a substantial spin
moment of 2.2
$\mu_{\text{B}}$, and W induces the W spin polarization. In the case of
Gd the spin polarization of the $d$ states is small (0.5 to 0.6
$\mu_{\text{B}}$) and this seems to be insufficient to induce spin
polarization via hybridization. This argumentation was checked by
performing calculations for Ni/W multilayers where Ni has roughly the
same spin polarization as the Gd $d$ band. 
Additional differences
between the calculations are that Fe/W, Gd/W, and Ni/W multilayers
have bcc, hcp, and fcc superstructures, respectively. Our calculations
found that Ni induced a much smaller spin moment of 0.005
$\mu_{\text{B}}$ in the W layers. The different crystal structures are
presumably responsible for inducing a spin moment for W on Ni but not
for W on Gd.


\subsection{Oscillatory behavior of W spin and orbital magnetic moments}


Whilst we have established conclusively that the spin magnetic moment 
on W is induced through hybridization, the oscillatory behaviour of 
the W spin magnetic moment needs more careful consideration. 
In Table I we compare charge-transfer modulations with the oscillation 
in the spin magnetic moment layer by layer. 
The charge-transfer modulation varies as $+-+-$ whilst the spin magnetic 
moment varies as $-++-$ and the orbital magnetic moment as $--++$. 
This means that Hund's third rule is violated in layers 1 and 3, 
whilst not in layers 2 and 4. 
This is a striking result emphasizing an oscillatory behaviour 
in the violation of Hund's third rule. 
Actually, this demonstrates that it is not very meaningful 
to discuss the behaviour of the spin and orbital magnetic moments 
in terms of Hund's third rule.

The striking feature of this work is the sign of the induced orbital
magnetic moment in the W interface layer. To determine the robustness
of this feature we plot in Fig.~\ref{filling} the spin and orbital
moments as a function of energy for a 5Fe(001)/3W multilayer. At the
energy position of $E_{\text{F}}$ ($E=0$) the proper value of the spin and orbital moments 
in the interface W layer are $-$0.113 and $-$0.031, respectively.
We see that for ``bulk''
Fe (Fe$_{3}$) the spin and orbital moments are aligned, the spin moment is
positive for all energies and the orbital moment changes sign around
$-0.05$ Ry. This would imply that if we reduce the number of electrons
in the system and in this manner model the lower Z systems, i.e.\
moving towards and lower than half filling, the spin and orbital moment
would become antiparallel aligned, consistent with Hund's third rule.
All Fe layers possess this behavior. In the W layers the spin moments
as a function of band filling show an oscillatory behavior, very
different from what was noticed in Fe. Moreover this oscillatory
behavior is very different between the W interface layer and the next W
layer. The orbital moments also oscillates as a function of band
filling and the simple cross-over from negative to positive values as
seen in Fe and shown to be consistent with Hund's third rule, does not
occur. Hence, the behavior can be seen to be a consequence of
hybridization between the
${\mathsf{5d}}$ and Fe and has no resemblance to Hund's third rule
behavior.


\subsection{Electronic and magnetic properties of 5Fe(001)/3Ta,
3W, and 3Re}

The spin and orbital moments as well as
the charge transfer for 5Fe(001)/3(${\mathsf{5d}}$), where
${\mathsf{5d}}$ is Ta, W, or Re, is displayed in Fig.~\ref{SLQ}.
The numerical values are given in Table~\ref{Qtab2}. Judging from the reduction of
the Fe spin moment at the interface, the hybridization of Ta with Fe is
strongest. Apart from this, the behavior of the spin moments as a
function of layers is similar for Ta, W, and Re. The figure shows that
the oscillations in the charge transfer do follow the oscillations in
the spin moments. However, concerning the orbital moments their behavior
varies from Ta to W to Re. For the 5Fe/3Ta there is no sign change in
the multilayer at all.
In the case of 5Fe/3Re the oscillation of the orbital
moment does follow the
 oscillations in the spin moments. For this system Hund's rule is not
violated either: in both Re layers the spin and orbital moments are
aligned. 
It is remarkable to see how complicated the behavior of the orbital
moment is as a function of atomic number and layer index, whilst we
note again the correlation between oscillations in the spin magnetic
moment and the charge oscillations.

This complicated behavior of the orbital moments in Ta, W, and Re and 
also as a function of layer index should not come anymore as a surprise 
given the complex behavior seen in Fig.~\ref{filling} as a function of energy or band filling. 
In particular by considering the energy scales in W1 and W2 of Fig.~\ref{filling} marked by a dashed box, 
we can see that for a reduced energy at the lower end of the box the orbital magnetic moment 
would become positive in both W1 and W2. In a rigid band fashion this would correspond to Ta. 
Increasing the energy towards the upper end of the box one can see that the orbital magnetic moment 
can remain negative in W1 but can become positive in W2.


\subsection{Comparison with calculations of ${\mathsf{5d}}$ impurities
in an Fe host}

The system that we consider is a three-layer system
incorporated in an Fe host. This three-layer system is repeated
periodically with the structure 5Fe(001)/3(${\mathsf{5d}}$). Whilst
such a geometry is far removed from a ${\mathsf{5d}}$ impurity in an
Fe host, comparison with calculations for such systems could be
instructive. Figure~\ref{5dseries} shows the spin and orbital moments
obtained for the 5Fe/3(${\mathsf{5d}}$) multilayer systems and the
${\mathsf{5d}}$ impurity calculations from Ref.~\onlinecite{Ebert}. The
spin moments of ${\mathsf{5d}}$ impurities in Fe are antiparallel
aligned for the first part of the
$5d$ transition metal series, i.e.\ for the ${\mathsf{5d}}$ with less
than half filling. From Os onwards the spin moments of the
${\mathsf{5d}}$ impurity and Fe become aligned. This is for the same
reasons as discussed in Sec.~\ref{sec:A}.
For ${\mathsf{5d}}$ impurities to obey Hund's third rule
the orbital moments would have to be positive throughout this
${\mathsf{5d}}$ series (c.f.\ Fig.~\ref{Hund} for the atomic case). This
is not the case and the orbital moments are negative for Re, Os, Ir; so
that in these calculations Hund's rule is violated for Os and Ir. The
spin and orbital moment curves for the multilayers show the same
behavior as for the impurity. For the spin moment the
multilayer and the impurity calculation switch between parallel and
antiparallel alignment at the same location. The orbital moments which
are more sensitive to the structure and chemical environment switch
between parallel and antiparallel at slightly different locations,
making the orbital moments negative for W, Re, and Os. Therefore,
Hund's rule is not only violated for W but also for Os as was also the
case for the single impurity.

Figure~\ref{filling} reinforces this universal band filling scenario.
When analyzing the orbital moment in the W interface layer (W$_{1}$) as
a function of  band filling in the vicinity of $E_{\text{F}}$ (the area
in Fig.~\ref{filling} marked by the box) we see that moving away from
$E_{\text{F}}$ to lower energy the orbital moment changes sign. Moving
from
$E_{\text{F}}$ to lower energy  in Fig.~\ref{filling} means, in a
rigid-band manner, mimicking lower Z such as Ta and therefore modelling
the Z lower than W in Fig.~\ref{5dseries}. Moving from $E_{\text{F}}$ up
in energy in Fig.~\ref{filling} also models the behavior of the orbital
moment in Fig.~\ref{5dseries}, but now for Z larger than W.
Therefore, we see that the behavior of the orbital moment as a
function of Z as shown in Fig.~\ref{5dseries} is present in the
behavior of the orbital magnetic moment of the W interface layer  as a
function of band filling (boxed area in Fig.~\ref{filling}).  Moreover,
a similar behavior can also be noted from Fig.~\ref{filling} in  the
layer W$_{2}$, adjacent to the interface layer. The difference
is that the area where the orbital moment is negative  has been
reduced. As a consequence Re, which is just after W in the  periodic
table and can be modelled in a rigid band fashion by moving up $E_{\text{F}}$,
has a negative orbital moment in the interface layer but
a  positive orbital moment in the adjacent W$_{2}$ layer
(Fig.~\ref{SLQ}). This implies that the energy has moved up in such a
manner that the orbital moment in the interface remains
negative but at the same time has moved up sufficiently to
reach energies where the sign change in the orbital moment of
the W$_{2}$ occurs ($\sim$0.05 Ry). Moving  $E_{\text{F}}$
down by roughly the same amount (0.05 Ry) to model Ta leads to positive
orbital moments in both the interface and the adjacent layer, in
agreement with the observed behavior of Fig.~\ref{SLQ}.


\subsection{Interface spin and orbital magnetic moments in the
${\mathsf{4d}}$}

If the band filling is the crucial ingredient in determining the
behavior of the orbital moments as a function of Z for the
${\mathsf{5d}}$ in Fe, then we would expect a similar behavior for the
${\mathsf{4d}}$ in Fe.  In Fig.~\ref{4dseries} we show the induced
spin and orbital magnetic moments for the ${\mathsf{4d}}$ obtained for
a 5Fe/3(${\mathsf{4d}}$) system. The spin moment is aligned
antiparallel to Fe in the first part of the transition metal series and
becomes parallel aligned from Rh onwards. Again this is a consequence of
hybridization as explained in Sec.~\ref{sec:A} and therefore the
${\mathsf{4d}}$ shows the same behavior as a function of band filling
as the
${\mathsf{5d}}$. Also the orbital magnetic moment shows the same
trends in the ${\mathsf{4d}}$ as in the
${\mathsf{5d}}$, namely from parallel to antiparallel
and again to parallel with respect to the Fe orbital moment with
zero crossings at Mo and Ru. These $4d$ elements are isoelectronic
with W and Os, and the zero crossings in the ${\mathsf{5d}}$ occur between
Ta and W and between Os and Ir. This shows that the zero crossings are only
slightly different in the ${\mathsf{4d}}$ and
${\mathsf{5d}}$.
In particular, the loci of zero orbital magnetic moment are shifted  
and the region of negative orbital magnetic moment
is larger for the $5d$ series than for the $4d$ series.
The loci of zero orbital magnetic moment for the ${\mathsf{4d}}$ impurity
occur at slightly higher values that for the ${\mathsf{4d}}$ interface.
This trend is consistent with the difference between impurity and interface of the 
${\mathsf{5d}}$ series as seen in Fig. \ref{5dseries}.
This indicates that the geometrical differences between impurity and interface alter
in a consistent way the loci of zero orbital magnetic moment to higher Z in the impurity
for both the ${\mathsf{4d}}$ and
${\mathsf{5d}}$ series.

For the
${\mathsf{5d}}$ impurity the zero crossings are between W and Re, and between
Ir and Pt. The similar behavior of these orbital moments as a
function of Z with only small changes between
the ${\mathsf{5d}}$ impurity and interface, and between
the ${\mathsf{4d}}$ impurity and interface demonstrates that band filling determines
these properties.


\section{Conclusions}

We have presented a systematic study of the interfacial spin and
orbital magnetic moments over the entire ${\mathsf{4d}}$ and
${\mathsf{5d}}$ series. So far only an extrapolation from two
experimental points, namely Ir and W, existed \cite{Wilhelm}. 

For W/Fe(001) systems with different W layer thicknesses we found that
in the Fe interface layer the Fe spin moment is reduced whilst its
orbital moment is approximately doubled. In the W layers a spin moment
is induced which is antiparallel to that of Fe in the first and fourth W layers but
parallel to Fe in the second and third W layers.  
The W orbital moment does not follow the spin moment. It is aligned
antiparallel to Fe in the first two W layers and changes sign in the
third and fourth W layers.
The calculations show
that small changes in the band filling can lead to a reversal of the
orbital moment of the first W layer. Hence, the behavior can be seen as
a consequence of the hybridization between W and Fe which is mainly due
to band filling, and to a lesser extent geometrical effects of either
single impurity or interface. 

In comparison with impurity calculations we found that the trend in
the spin and orbital moments over the $4d$ and $5d$ series is quite similar. The
number of zero crossings is the same for the interface and impurity
calculations, again suggesting that the behavior is primarily due to
band filling. Furthermore, for the ${\mathsf{4d}}$ interface we find
the same behavior as for the ${\mathsf{5d}}$, despite the
fact that the spin-orbit interaction of the $4d$ electrons is much
smaller than for $5d$ electrons.

\medskip

{\bf Acknowledgements.}

This work has been partially funded by the EU Research
Training Network in ``Computational Magnetoelectronics''
(HPRN-CT-2000-00 143).


\begin{figure}    
\caption{
Atomic Hund's rule values for the spin moments (full line) and
orbital magnetic moments (dashed line) across the transition metal
$d$ series.}  
\label{Hund}
\end{figure}

\begin{figure}    
\caption{Spin moments (top panel) and orbital moments (lower panel)
in units of $\mu_{\text{B}}$ for the different layers of the
7Fe(001)/$m$W multilayer with $m$=
$\{$1,3,5,7$\}$.}  
\label{LSthickness}
\end{figure}

\begin{figure}
\caption{Centroid positions of spin-up ($\bigtriangleup$) and spin-down
($\bigtriangledown$) bands for the different layers of the 5Fe/3W
system. The inset shows schematically the spin magnetic moments in the layers of this
system.}
\label{Centroid}
\end{figure}

\begin{figure}
\caption{Layer-resolved density of states for the 5Fe/5W system;
separated in majority spin (full line) and minority spin
(dashed line) bands.}
\label{DOS}
\end{figure}

\begin{figure}
\caption{Spin moments (full line) and orbital moments (dashed line)
in units of $\mu_{\text{B}}$ for the 5Fe/3W system as a function of energy.
The vertical dashed line corresponds to the Fermi energy. For clarity
the orbital moments in all layers are $\times$10 and the
spin moments in the W layers are $\times$5.} 
\label{filling}
\end{figure}

\begin{figure}
\caption{Spin moments (top panel) and orbital moments (middle panel) (in
units of $\mu_{\text{B}}$) and
charge transfer (lower panel) for the layers in the systems
5Fe(001)/3(${\mathsf{5d}}$) with ${\mathsf{5d}}$=$\{$Ta, W, Re$\}$. A
negative charge transfer corresponds to a charge excess on that sphere.}
\label{SLQ}
\end{figure}

\begin{figure}
\caption{Comparision of the theoretical spin moment (top panel) and
orbital moment (lower panel) between the 5Fe(001)/3(${\mathsf{5d}}$)
multilayer system (solid line) and the ${\mathsf{5d}}$ impurities in Fe
[Ref.~\onlinecite{Ebert}] (dashed line) across the $5d$ transition metal
series.}
\label{5dseries}
\end{figure} 

\begin{figure}
\caption{Comparision of the theoretical spin moment (top panel) and
orbital moment (lower panel) between the 5Fe(001)/3(${\mathsf{4d}}$)
multilayer system (solid line) and the ${\mathsf{4d}}$ impurities in Fe
[Ref.~\onlinecite{Ebert}] (dashed line) across the $4d$ transition metal
series.}
\label{4dseries}
\end{figure}


\begin{table}
\caption{Layer-resolved charge transfer ($\Delta Q$), $d$ and $sp$
occupancy of the majority and minority spin channel, and spin and
orbital magnetic moments for 5Fe(001)/$m$W multilayer systems with
$m$=$\{$1,3,5,7$\}$.}
\label{Qtab1}
\begin{center}
\begin{tabular}{cccccccc}
\hline
{\bf{system}}  & charge & \multicolumn{2}{c}{$d$ occ.} &
\multicolumn{2}{c}{$sp$ occ.} & \multicolumn{2}{c}{moments} \\ 
layer & $\Delta Q$  & up & down & up & down  &
$\mu_S$  & $\mu_L$  \\
\hline
{\bf{5Fe/W}} &&&&&&& \\
Fe3   & 0.011 &  4.404 & 2.113 & 0.703 & 0.769  & 2.225 & 0.043 \\
Fe2   & 0.001 &  4.455 & 2.049 & 0.718 & 0.777  & 2.347 & 0.051 \\
Fe1   & $-0.134$ &  4.266 & 2.347 & 0.743 & 0.778  & 1.884 & 0.085 \\
W1    & 0.256 & 1.992 & 2.185 & 0.745 & 0.822  & $-0.270 $ & $-0.041 $ \\
{\bf{5Fe/3W}} &&&&&&& \\
Fe3   & 0.006 & 4.346 & 2.162  & 0.709 & 0.777  & 2.116 & 0.039 \\
Fe2   & 0.005 & 4.421 & 2.059  & 0.729 & 0.786  & 2.305 & 0.043 \\
Fe1   & $-0.119$ & 4.161 & 2.415  & 0.752 & 0.791  & 1.707 & 0.089 \\ 
W1    & 0.137 & 2.087 & 2.162  & 0.788 & 0.826  & $-0.113 $ & $-0.031$ \\
W2    & $-0.055$ & 2.226 & 2.183  & 0.824 & 0.822  & 0.045 & $-0.041$ \\
{\bf{5Fe/5W}} &&&&&&& \\
Fe3   &  0.007 & 4.301 & 2.190   & 0.719 & 0.784  & 2.046 & 0.033 \\
Fe2   & 0.005 &  4.393  & 2.069   & 0.738 & 0.794  & 2.267 & 0.041 \\
Fe1   & $-0.121$ &  4.136 & 2.425  & 0.759 & 0.800  & 1.670 & 0.078 \\
W1   & 0.138 &  2.077 &   2.151  & 0.797 & 0.836  & $-0.113$ & $-0.023$ \\
W2   & $-0.029$ &  2.198  & 2.166   & 0.834 & 0.831  & 0.035 & $-0.016$ \\
W3   & 0.007 &  2.193 & 2.141  & 0.830 & 0.828  & 0.054 & 0.005 \\
{\bf{5Fe/7W}} &&&&&&& \\
Fe3   &  0.007 &  4.285 & 2.192  & 0.725 & 0.791  & 2.027 & 0.032 \\
Fe2   & 0.005 &  4.374 & 2.076  & 0.744 & 0.802  & 2.240 & 0.039 \\
Fe1   & $-0.121$  &  4.123 & 2.425  & 0.766 & 0.808  & 1.656 & 0.076 \\
W1    & $ 0.138$ &  2.074 & 2.138  & 0.804 & 0.844  & $-0.105$ & $-0.022$ \\
W2    & $-0.029$ &  2.194 &  2.155  & 0.841 & 0.839  & 0.042 & $-0.017$ \\
W3    & 0.004 &  2.174 &  2.147  & 0.839 & 0.836  &  0.031 & 0.001 \\
W4    & $-0.001$ &  2.156 &  2.166  & 0.842 & 0.837  & $-0.005 $ & 0.005 \\
\hline
\end{tabular}
\end{center}
\end{table}


\begin{table}
\caption{Layer-resolved charge transfer ($\Delta Q$), $d$ and $sp$
occupancy of the majority and minority spin channel, and spin and
orbital magnetic moments for 5Fe(001)/3(${\mathsf{5d}}$)
multilayer systems with
${\mathsf{5d}}$=$\{$Ta, W, Re$\}$.}
\label{Qtab2}
\begin{center}
\begin{tabular}{cccccccc}
\hline
{\bf{system}}  & charge & \multicolumn{2}{c}{$d$ occ.} &
\multicolumn{2}{c}{$sp$ occ.} & \multicolumn{2}{c}{moments} \\
 layer & $\Delta Q$  & up & down & up & down  &
$\mu_S$  & $\mu_L$  \\
\hline
{\bf{5Fe/3Ta}} &&&&&&& \\
Fe3   &  0.014 & 4.239 & 2.227  & 0.729 & 0.791  & 1.950 &  0.043 \\
Fe2   & $-0.013$ & 4.301 & 2.161  & 0.745 & 0.806  & 2.079 &  0.042 \\
Fe1   & $-0.078$ & 4.067 & 2.503  & 0.728 & 0.780  & 1.512 &  0.057 \\
Ta1   &  0.105 & 1.568 & 1.731  & 0.774 & 0.822  & $-0.209$ &  0.044 \\
Ta2   & $-0.040$ & 1.697 & 1.755  & 0.788 & 0.800  &  $-0.070$ &  0.034 \\
{\bf{5Fe/3W}} &&&&&&& \\
Fe3   & 0.006 & 4.346 & 2.162  & 0.709 & 0.777  & 2.116 & 0.039 \\
Fe2   & 0.005 & 4.421 & 2.059  & 0.729 & 0.786  & 2.305 & 0.043 \\
Fe1   & $-0.119$ & 4.161 & 2.415  & 0.752 & 0.791  & 1.707 & 0.089 \\ 
W1    & 0.137 & 2.087 & 2.162  & 0.788 & 0.826  & $-0.113 $ & $-0.031$ \\
W2    & $-0.055$ & 2.226 & 2.183  & 0.824 & 0.822  & 0.045 & $-0.041$ \\
{\bf{5Fe/3Re}} &&&&&&& \\
Fe3   & 0.005 & 4.397 & 2.125  & 0.701 & 0.772  &  2.201 &  0.038 \\
Fe2   & 0.009 & 4.451 & 2.047  & 0.716 & 0.777  &  2.343 &  0.043 \\
Fe1   & $-0.129$ & 4.274 & 2.298  & 0.762 & 0.795  & 1.943 &  0.095 \\
Re1   & 0.144 & 2.594 & 2.631   & 0.796 & 0.835  &  $-0.076$ & $-0.043$ \\
Re2   & $-0.055$ & 2.738 & 2.637  & 0.841 & 0.839  & 0.103 &  0.022 \\
\hline
\end{tabular}
\end{center}
\end{table}


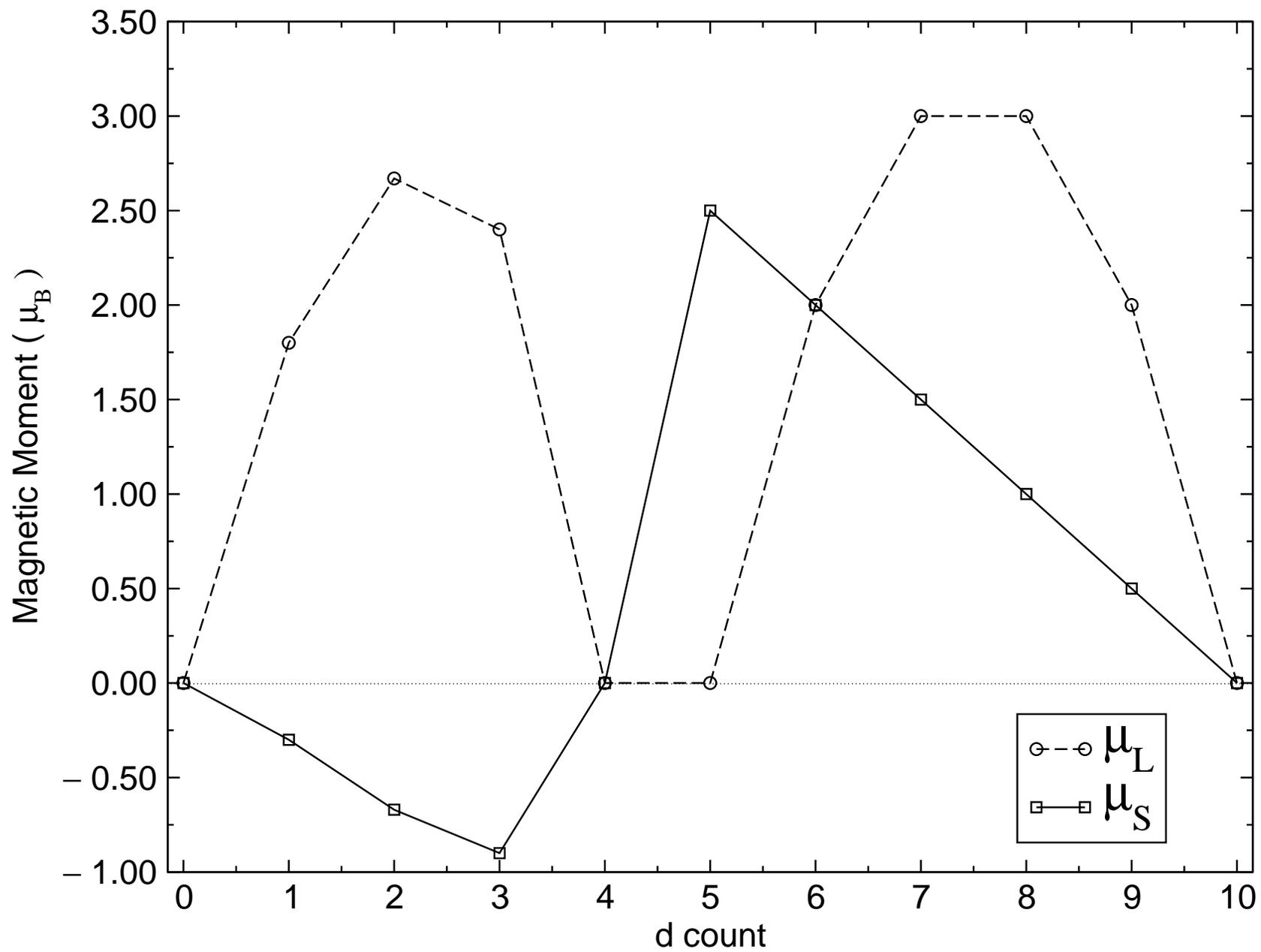
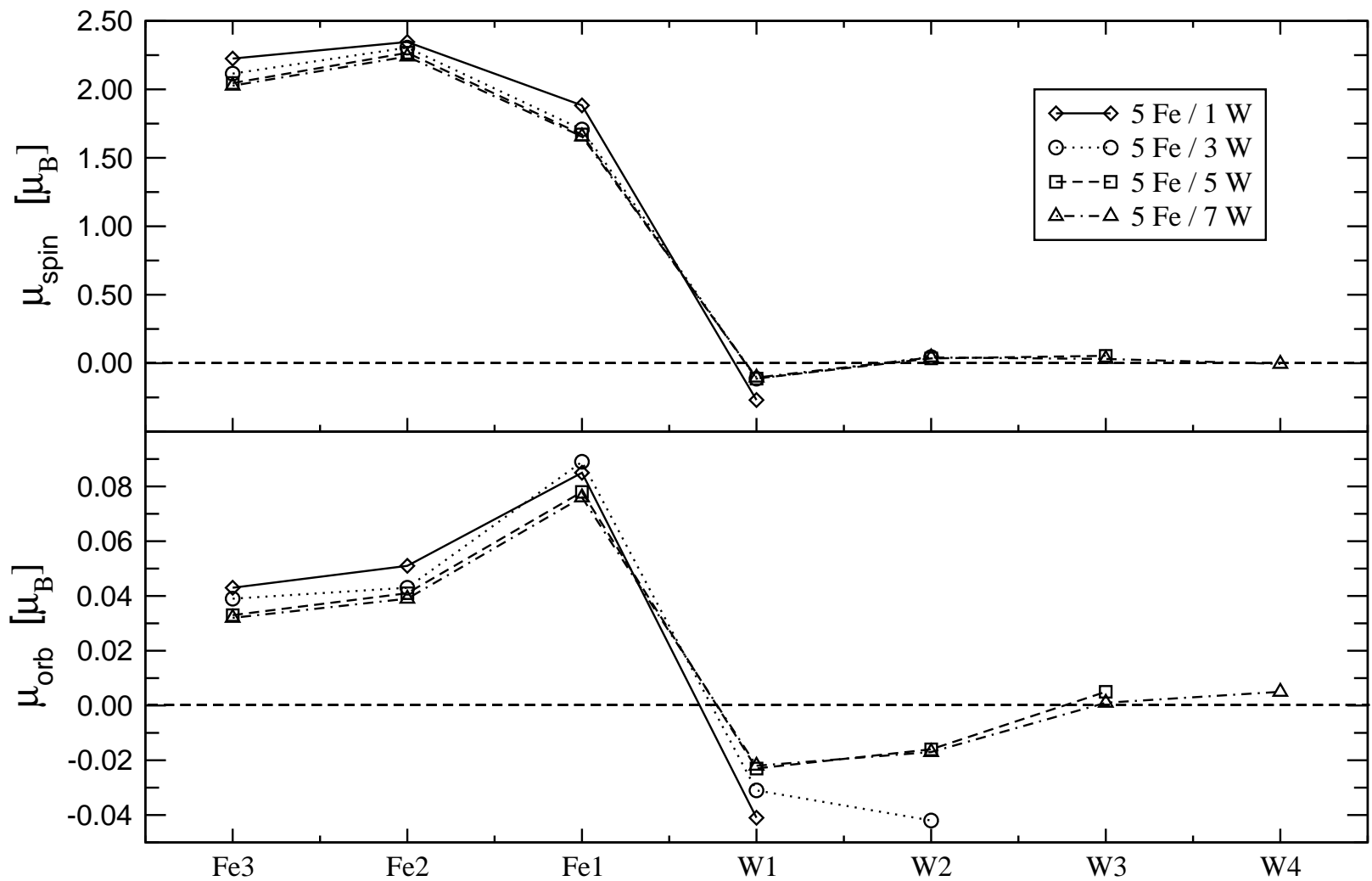
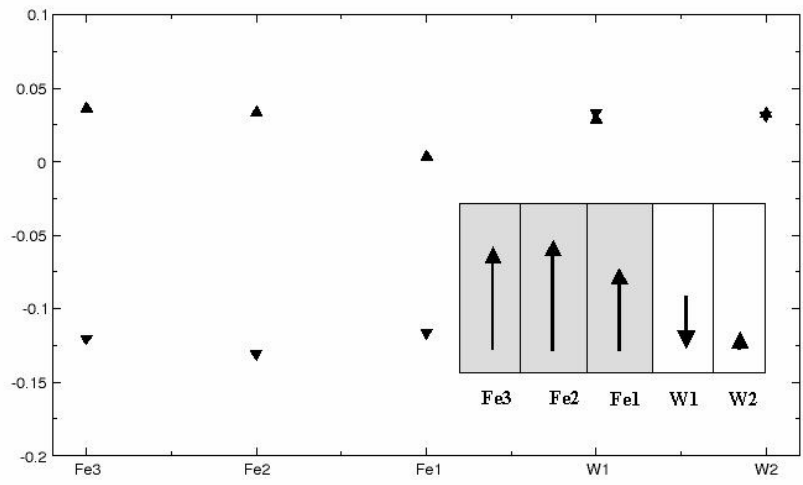
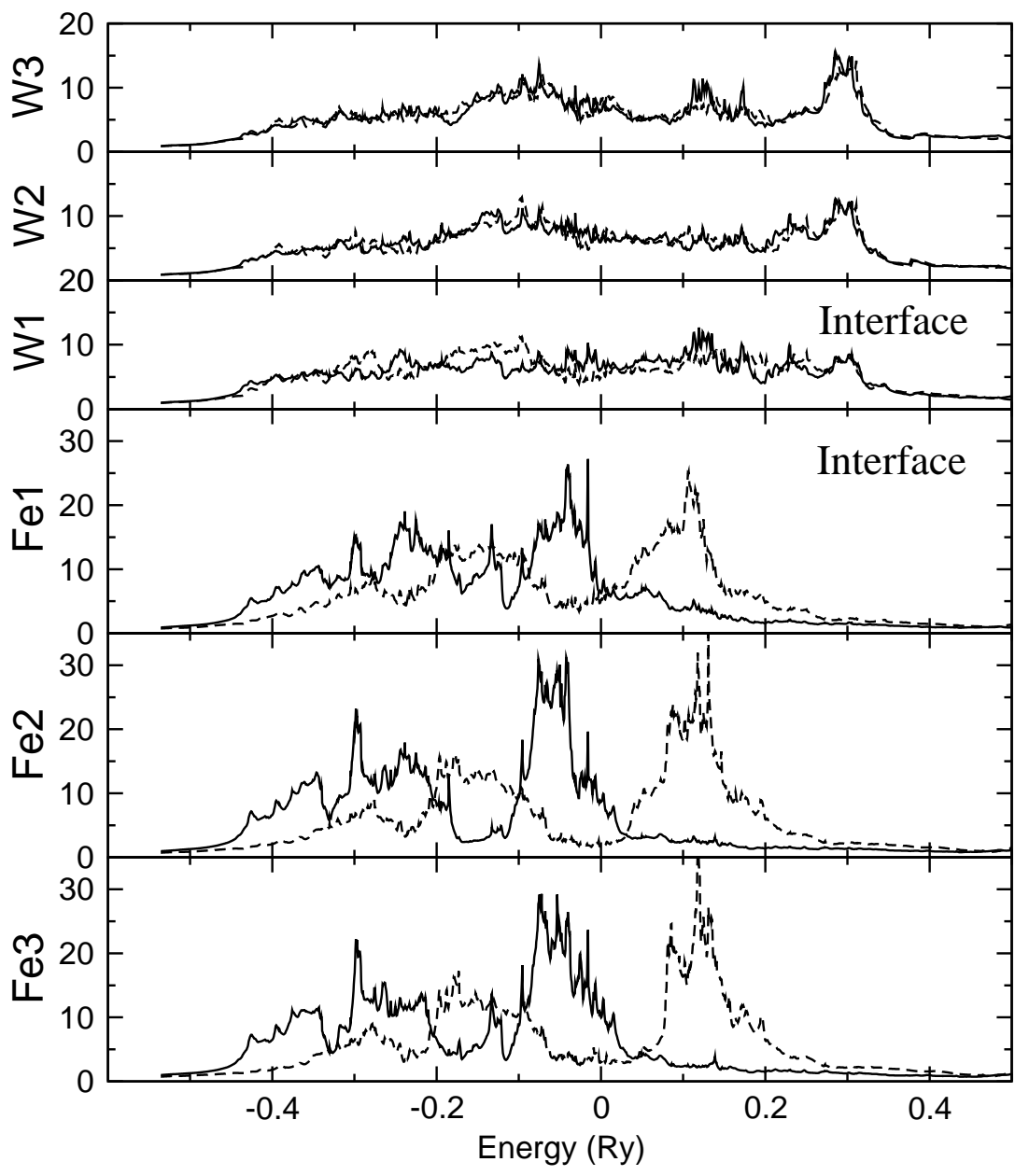
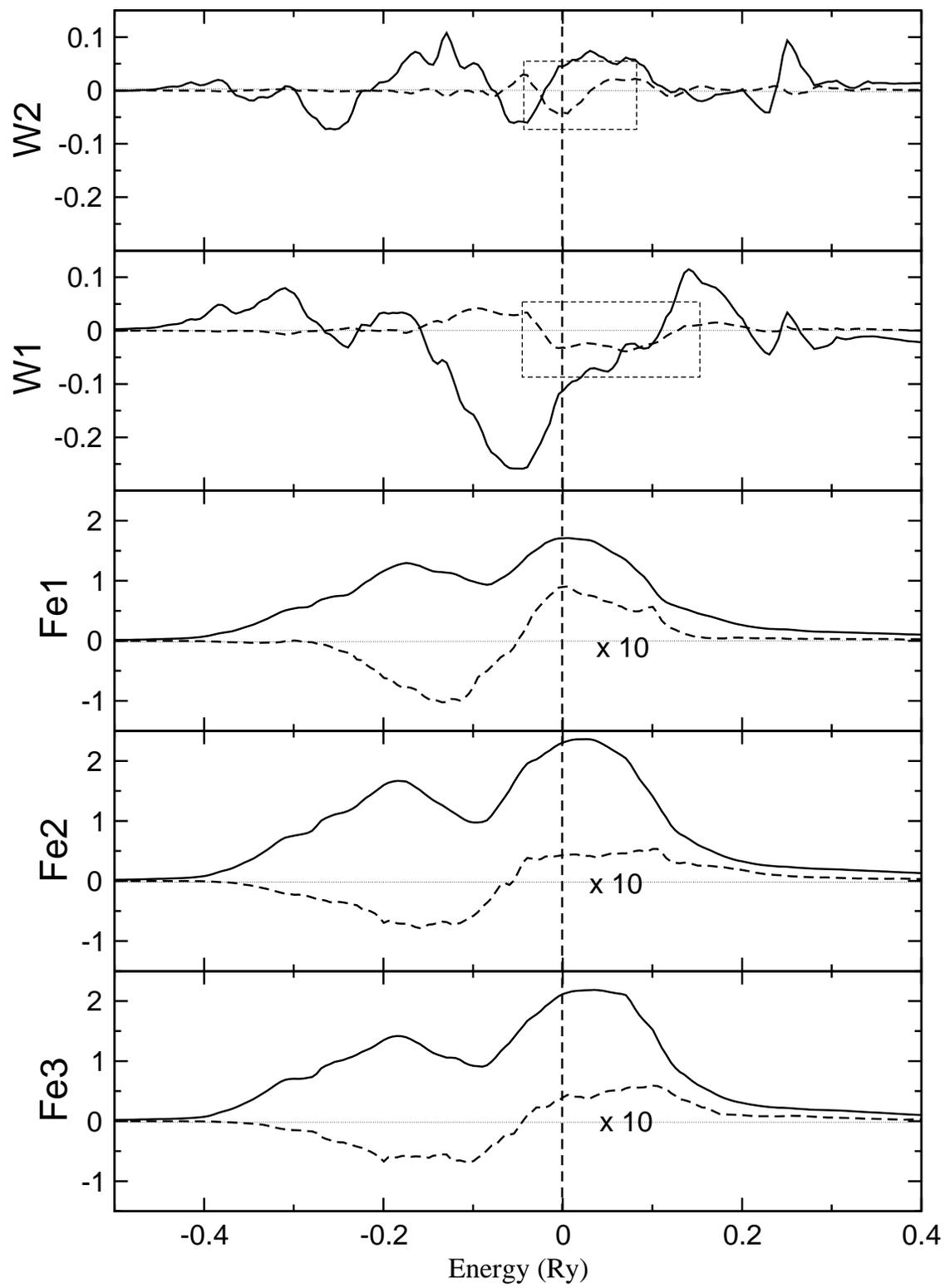
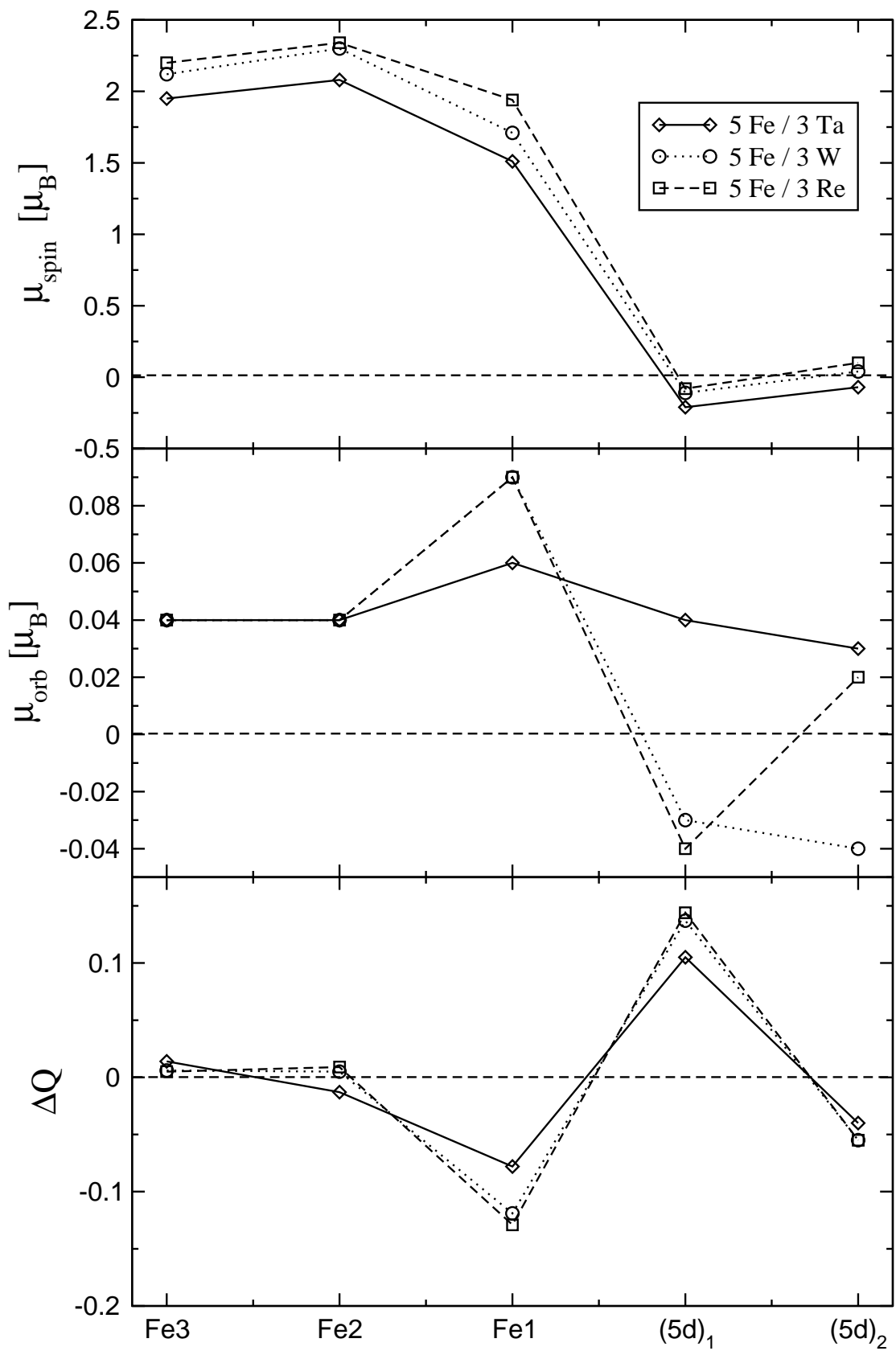
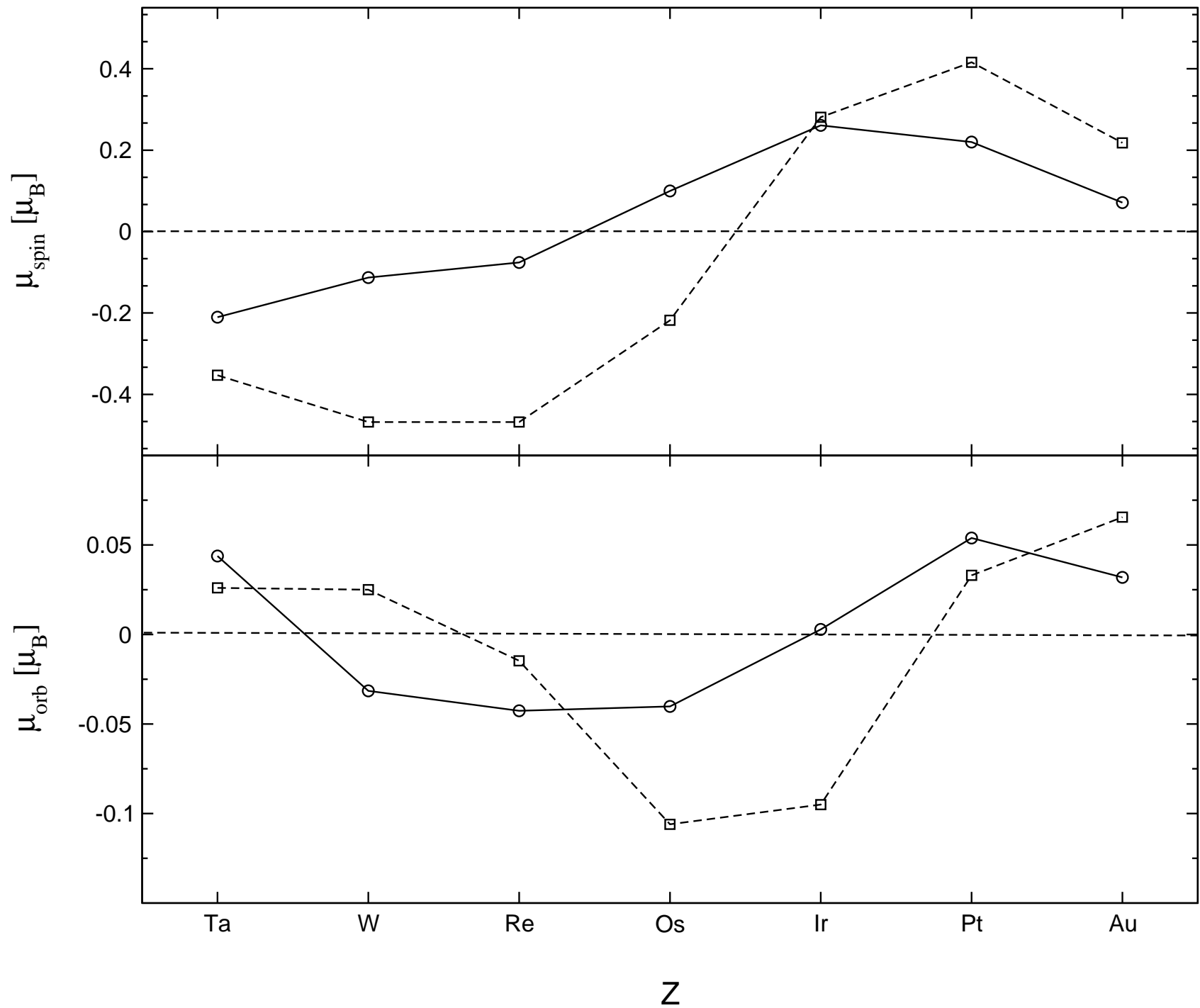
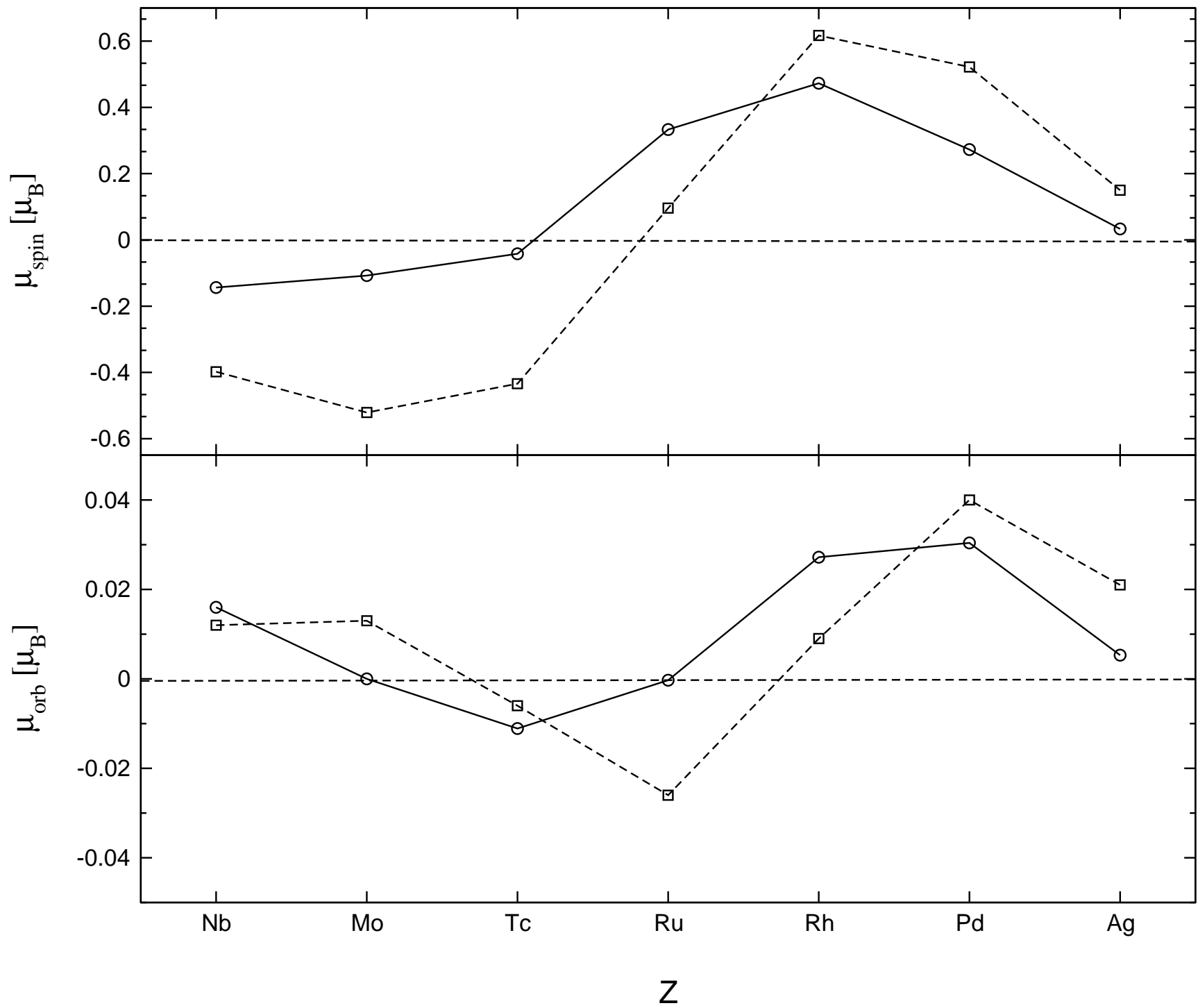
\end{document}